\documentclass{article}
\usepackage{graphicx} % Required for inserting images

\title{Deep learning assisted SERS detection of prolines and hydroxylated prolines using nitrilotriacetic acid functionalized gold nanopillars}
\author{Yuan Zhang a,Kuo Zhan a,b,  Peilin Xin a,b, Yingqi Zhao a,b,}

\date{Shubo Wang d, Aliaksandr Hubarevich e, Xuejin Zhang f,
Jianan Huang a,b,c, *.
Affiliation: 
a Research Unit of Health Sciences and Technology (HST), Faculty of Medicine, University of Oulu, Oulu 90220, Finland.
 b Biocenter Oulu, University of Oulu, Oulu 90220, Finland.
c Research Unit of Disease Networks, Faculty of Biochemistry and Molecular Medicine, University of Oulu, Oulu 90220, Finland.
d Nano and Molecular Systems Research Unit, Faculty of Science, University of Oulu, Oulu 90570, Finland.
e Plasmon Nanotechology Unit, Istituto Italiano di Tecnologia, Via Morego 30, 16163 Genova, Italy
f School of Physics and College of Engineering and Applied Sciences, Nanjing University, Nanjing 210093, China
* Corresponding author. Email: jianan.huang@oulu.fi}

\begin{document}

\maketitle

\section{Abstract}
Abstract 
Proline (Pro) is one kind of proteinogenic amino acid and an important signaling molecule in the process of metabolism. Hydroxyproline (Hyp) is a product on Pro oxygen sensing post-translational modification (PTM), which is efficiently modulated tumor cells for angiogenesis. Distinguishing between Pro and Hyp is crucial for diagnosing connective tissue disorders, as elevated levels of Hyp can indicate abnormal collagen metabolism, often associated with diseases like osteogenesis imperfecta or fibrosis. However, there is a very small difference (hydroxyl group) between molecular structures of Pro and Hyp, which is a big challenge for current detection technologies to distinguish them. For surface-enhanced Raman scattering (SERS) sensors, the similar molecule structure leads to similar Raman spectra that are difficult to distinguish. Furthermore, another problem is the weak affinity between amino acids sample and SERS-active substrates by physical adsorption. The selecting capturing of Pro and Hyp in the mixture of amino acids is not easy to achieve. In this work, we designed a new method for Pro and Hyp specifical detection and recognition by using gold nanopillars as the SERS substrate and combing nitrilotriacetic acid (NTA) with nickel (Ni) to form NTA-Ni structure as a specifical affinity agent. One side of NTA-Ni was attached to gold nanopillars through thiol binding. Another side captured the amino acids using reversible binding by receptor-ligand interaction between Ni and amino acids. Because of the different binding time with NTA-Ni and amino acids, the sensor can recognize Pro and Hyp from amino acids mixture. Then we used automatic peak assignment program for data analysis and machine learning model to distinguish between Pro and Hyp. The label-free SERS detection of amino acids PTM using gold nanopillars provides a potential method to further biomolecule detection and specifical capture.

\section{Introduction}

Proline (Pro) is an important proteinogenic amino acid, which has an exceptional structure and is fundamental for many metabolic process.1 Hydroxyproline (Hyp) is a non-essential amino acid derived from Pro through a oxygen sensing post-translational modification (PTM) that is dynamically regulated tumor cell metastasis.2-7 The modification is happened by prolyl hydroxylases that add a single hydroxyl group (-OH) to replace the hydrogen of the Pro pyrrolidine ring. Hyp can subtly influence the protein structures, activities, and the properties of protein-protein interactions in the cell and play a pivotal role in cancer development and disease progression.8 Therefore, distinguishing Hyp from Pro is crucial for studying collagen structure, monitoring diseases related to collagen metabolism, analyzing metabolic pathways, and evaluating therapeutic or nutritional outcomes.9, 10

Current methods for detection of Pro and Hyp have their limitations. Isatin paper assays are simple but not sensitive.11 Colorimetric techniques, like ninhydrin reactions, are reliable but influenced by reaction variability.12 High-performance liquid chromatography (HPLC) provides detailed amino acid profiles but is costly and time intensive.13 Mass spectrometry offers high sensitivity and detailed modifications but requires expertise and expensive equipment.14 Alternatively, surface-enhanced Raman spectroscopy (SERS) is promising for detection of Pro and Hyp, because it combines the molecular specificity of vibrational Raman spectroscopy with high sensitivity to provide rich fingerprint information of analytes.  

However, the small difference of molecular structure (-OH group) between the Hyp and Pro is challenging for reproducible SERS detection of Pro and Hyp, which required long physical adsorption time of the analytes on metal colloids. For example, J. J. Cárcamo et al. 15 prepared silver colloids at a controlled and fixed pH, allowing the analyte to be physically adsorbed on the colloid surface to obtain unique and reproducible SERS spectra. Accordingly, the required adsorption time for Pro and Hyp in the silver colloids for distinguishable and reproducible SERS spectra were 72 hours for Pro and 48 hours for Hyp. Similarly, Ariel R. Guerrero et al. employed gold colloids instead of the silver colloids to obtain reproducible SERS spectra of Hyp with distinctive bands.16 These work implied that physical adsorption would be difficult for practical SERS detection of mixture of Hyp and Pro in biofluid, because the colloid surface would be mostly occupied by Hyp.  

Recently, Zichen Yang et al. functionalized silver colloids with covalent organic frame with carboxylate groups (Ag@COF-COOH) to capture the Hyp based on the hydrogen binding mechanism for SERS detection. 17 The detection time was far less than that of analyte adsorption on colloids, which was then used in fast Hyp detection in the inferior milk. However, the authors did not demonstrate SERS detection of Pro. Meanwhile, the Ag@COF-COOH has the same binding probability for both Hyp and Pro, which means that it could not distinguish Hyp from Pro. Because Pro and Hyp co-exist in many biological fluids, the accuracy of using Ag@COF-COOH to detect Hyp in milk might be problematic. Therefore, to achieve distinguishing Hyp from Pro in biofluids by SERS, an effective affinity agent to bind uniquely both analytes on plasmonic nanostructures is necessary to build a database of reproducible and distinguishable SERS spectra of Hyp and Pro.

In this study, we bound nitrilotriacetic acid (NTA) with nickel (Ni) to form NTA-Ni structure as the affinity agent on plasmonic gold nanopillars (AuNPs) for label-free and rapid SERS discrimination of Pro and Hyp in liquid with a machine learning model. As conceptualized in Figure 1, the NTA was covalently attached on the surface of gold nanopillars via thiol bonding, then reacted with Ni to form NTA-Ni as the specific affinity agent to capture Pro and Hyp as NTA-Ni-Pro and NTA-Ni-Hyp through chelation.18,19 The gold nanopillar has the advantage of a wide-range hot spot with a decay length around 50 nm for reproducible SERS detection of large and long molecule moieties.20,21 As the length of NTA-Ni-Pro (NTA-Ni-Hyp) is around 4 nm,22 the hot spot of the gold nanopillars can cover the whole molecule chains to provide distinguishable SERS spectra. Combining machine learning analysis, our method is promising to analyze complex and similar molecules in biofluids towards rapid analysis and early diagnosis of diseases.

\section{Experimental section}

• Materials 

Silicon wafers 4-inch prime CZ-Si wafer, thickness = 525 ± 25 um, (111), 1-side polished, p-type (Boron), 1 - 10 Ohm cm was purchased from Micro chemicals; polystyrene with 0.5 um diameter was purchased from Micro Particles; ethanol of purity <99.8%, AnalaR NORMAPUR® ACS, analytical reagent, Nickel(II) chloride (NiCl2) were purchased from Fisher Scientific; ethylene glycol (HOCH2CH2OH), acetone (CH3COCH3), Sodium dodecyl sulfate solution (SDS, CH3(CH2)11OSO3Na), Hydrogen peroxide solution (H2O2), Hydrofluoric acid (HF), N-Nα,Nα-Bis(carboxymethyl)-L-lysine-12-mercaptododecanamide (NTA, C22H40N2O7S), L-Proline (C5H9NO2), trans-4-Hydroxy-L-proline (C5H9NO3) were purchase from Sigma-Aldrich.

•	Gold nanopillars fabrication

Gold coated silicon pillar fabrication was based on the established nanosphere lithography protocol.20 The procedures are following these steps, firstly, depositing 500 nm polystyrene nanospheres on the surface of 1x1 cm silicon wafers. The silicon wafers were hydrophilic by Oxygen plasma (Gambetti Tecnologia colibri and bench-top plasma system, Italy), an automatic pump was used to control the speed of polystyrene sphere droplet, when polystyrene sphere occupied full of petri dish, the cleaned wafers were used to carry out the polystyrene sphere single layer. Allow the wafers with polystyrene sphere to be naturally dry, then reactive ion etching (RIE, Oxford Instruments Plasmalab 80 Plus PECVD/ICP-DRIE, UK) instrument was used to reduce the diameter of polystyrene nanospheres from 500 nm to 250 nm. After that, sputter coating (Quorumtech Q150T ES, UK) 5 nm silver and 10 nm gold on the surface of silicon wafers. Then ultrasonication for 5 minutes to remove polystyrene nanospheres. 20 mL H2O, 1 mL HF and 0.2 mL H2O2 mixture solution was prepared for chemical etching of nanopillars. Finally, sputter coating 2 nm of titanium and 50 nm gold for the nanopillars, then get the complete gold nanopillars chips. Electronic microscopic images of the nanopillars were taken by Scanning Electron Microscopy (SEM, ZEISS Sigma HD VP FE-SEM, Germany). The reflectance of the as-made nanopillar chips was measured by in the integrating sphere of the Ultraviolet–visible spectroscopy (UV-Vis, Shimadzu UV- 2600, UK). X-ray Photoelectron Spectroscopy (XPS, Thermo Fisher Scientific ESCALAB 250Xi XPS, USA) was used to study the surface binding mechanism.

•	Samples preparation

Gold nanopillars hydrophilicity: To clean the gold nanopillars, they were rinsed in H2O2, ethanol, and H2O. Then, an oxygen plasma treatment (Gambetti Tecnologia colibri and bench-top plasma system, Italy) was performed to make them hydrophilic. The parameters of oxygen plasma are 50 mW power, 30 s, three times. 
NTA coating gold nanopillars: 2 mL 5 mM NTA solution was added into 60x15 mm petri dish, then put gold nanopillars chips into this petri dish for 24-48h, got the gold nanopillars with NTA (Au-NTA) chips. 
Raman measurement samples: Ethanol and H2O were used to clean the extra NTA on the surface of AuNPs-NTA chips. Then, 2 mL 5 mM NiCl2 was added to react with NTA for 30 min to form an AuNPs-NTA-Ni structure. After that, 2 mL 100 nM Pro or 100 nM Hyp neutral solution was directly added to the mixture for Raman measurements. 

•	SERS measurements

Raman spectroscopy was performed using a Thermo Scientific DXR2xi Raman Imaging Microscope. The working parameters were a laser wavelength of 785 nm, an expose time of 0.1-1 s (scanned once), a laser power of 2.5-10 mW, and a 60x water-immersion objective. None of the SERS spectra used any smoothing operation except for background removal. Finally, Andor Solis software was used to collect the data, it collected 2000 spectra in every data at the same time. every spectrum has. Raman data analysis was performed using the MATLAB R2022a and Origin 2022b software.

•	Data pre-processed and peak assignment

Peak assignments: MATLAB was used to do peak assignments and draw SERS spectral figures. The main steps are: (1) Plot the SERS spectra of 2000 raw data; (2) Cosmic ray removed: remove the influence from Ramam instrument; (3) Normalization: make all spectra have the same baseline; (4) Background removal: reduce the other purities influences; (5) Signal detection: collect all spectra with good signal; (6) Find peaks: use Pro and Hyp database to find their peaks in spectra; (7) Output figures results.
Distribution histograms of peak occurring frequency: MATLAB scripts were used to calculate the peak occurring frequency at each Raman shift. These steps are: (1) Pretreatment of multiple data; (2) Extracting effective spectra: find all spectra with rich signals; (3) Find peak frequency: summarize each peak’s frequency in all spectra; (4) Output Excel results, (5) Draw histograms Origin software was used to draw the distribution histograms of SERS peaks. 

•	Machine learning classification

Machine learning model: MATLAB was used to run the codes of convolutional neural network (CNN) model for classification and post-evaluation. A total of 6000 and 2000 peak assignment selected spectra with 1463 features were classified using 5-fold cross-validation for CNN classification model and post-evaluation model, where 80% of the spectra were used as the training set and 20% of the spectra were used as the test set. Following these steps: 1) Dataset Preparation, 2) Training Process, 3) Evaluation Metrics. 

The detailed codes are seen in Github link: KuoZHAN/Surface-enhanced-Raman-spectroscopic-detection-of-proline-and-hydroxylated-proline-: CNN and post-evaluation model with raw data.

\section{Results and discussion}

We firstly confirmed that wide-range hot spot of the AuNPs by the finite-difference time-domain (FDTD) simulation and reflectance measurement by UV-Vis spectrometry. According to the FDTD simulation result (Figure 2a), the AuNPs with diameter of 250 nm have the decay length is 50 nm at 785 nm Raman laser wavelength when the simulated reflectance shown a LSPR at 790 nm (Figure S1a). Accordingly, we fabricated AuNPs with similar diameter, the height of around 1 µm, the period of around 500 nm (Figure 2b, c). The measured reflectance shows that its LSPR location at 786 nm (Figure S1b). The fabricated AuNPs have uniform surface and stable structure (Figure S2). These results demonstrate that the fabricated AuNPs can provide reproducible SERS spectra with wide-range hot spot and high enhancement effect.

To test the binding capability of NTA to capture the Pro or Hyp, we measured time series SERS spectra of molecules in different systems: AuNPs (Figure S3a), AuNPs-Pro (Figure S3b), AuNPs-NTA (Figure 3a) and AuNPs-NTA-Ni-Pro (Figure 3b). After the NTA was incubated on the AuNPs for 24-48 hours to form AuNPs-NTA structure (see Experiment for details), the SERS time-series exhibited many NTA peaks in Figure 3a. By comparing the Raman spectrum of NTA powder (Figure S4), the continuous SERS peaks of NTA are found during 700-1300s: deformation of COOH chemical group (629 cm-1), bending of C=O (722 cm-1), stretching of C-C and CH2 (821 cm-1), bending of C-N (992 cm-1), wagging of CH2 (1133 cm-1), bending and twisting of CH2 (1208 cm-1), deformation of CH2 (1420 cm-1), respectively, as listed in Table S1.23-28

In the case of using AuNPs-NTA-Ni to bind Pro molecules, the SERS time-series spectra show overlapping peaks of both NTA and Pro as they have similar molecular structures (Figure 3b). For example, NTA has three COOH groups, while Pro has one. So, the symmetrical stretching peak at 919 cm-1 and the deformation peak at 809 cm-1, are the common features of both NTA and Pro. However, the unique peaks at 411, 618 cm-1 are from aromatic ring bending, and 1225 cm-1 as the Pro ring deformation of CH2 and NH2 (red arrow indicate in Figure 3b) demonstrates the successful binding of Pro. The 618 cm-1 shift to 648 cm-1 with the changes of Pro confirmation. Other peaks assignments information can be found in Table S2, and more SERS spectra are in Figure S5-7.

As a control sample without any molecule’s adsorption, no SERS signals from AuNPs themselves were observed (Figure S3a). When the Pro solution was directly added into the petri dish containing the AuNPs chips, the water-immersion objective was dipped into it immediately for SERS measurement. Such physical adsorption of Pro on the AuNPs for a short time also exhibited no SERS signals (Figure S3b).  

We further used XPS to confirm the molecule binding mechanisms in Figure 3c, d. The XPS survey spectra of AuNPs-NTA (Figure 3c) and AuNPs-NTA-Ni-Pro (Figure 3d) shared many intense AuNPs peaks (Au 4f, 83 eV; 4d, 335 eV; 4p, 546 eV), the presence of features corresponding to sulfur (2p, 162 eV; 2s, 227 eV), hydrocarbon (284 eV), nitrogen (1s, 399 eV; 1136 eV), oxygen (2s, 27 eV ;1s, 532 eV; 976 eV), which are similar with those in the reference.29 There are some bands from silver (3d, 368 eV; 573 eV; 605 eV), because the fabrication of AuNPs coated silver and gold together for the chemical etching step. Comparing the XPS peaks of AuNPs-NTA, a unique Ni (2p, 856 eV) peak is found, the oxygen (1s, 532 eV) peak is much higher, and the gold (4f, 83 eV) peak is much weaker in the AuNPs-NTA-Ni-Pro spectrum. They confirmed the binding of Ni and Pro that increased the oxygen signals and screened the gold signals. A close examination of high-resolution data for the S 2p core levels (Figure S8) showed the S2- double between 161 and 165 eV, which is consistent with literature reported data of Au-NTA.30 Other SERS spectra and XPS survey spectra of Hyp are shown in Figure S9. In addition to the SERS spectra, these XPS results provide additional evidence to prove the NTA molecules were bonded to the surface of AuNPs successfully and the formation of AuNPs-NTA-Ni-Pro. 

To classify the Pro and Hyp, we developed a convolutional neural network (CNN) model for the spectra classification of AuNPs-NTA-Ni-Pro and AuNPs-NTA-Ni-Hyp in Figure 4. Figure S10a shows the low loss curve and high accuracy curve of the training model as the increasing of the training epochs, which means the great classification performance of the trained CNN model for recognition of AuNPs-NTA-Ni-Pro and AuNPs-NTA-Ni-Hyp’s spectra. There are 4800 spectra data of AuNPs-NTA-Ni-Pro and AuNPs-NTA-Ni-Hyp trained in this CNN model, the classification accuracies are 99.5 % for Pro and 99.1 % for Hyp, as presented in Figure S10b.  The confusion matrix for the Pro and Hyp SERS classification is shown in Figure 4 step 1, with the accuracy of 96.5% to Hyp and 95.9 % to Pro. To test the classification stability of the trained CNN model, another 2000 SERS spectra data collected in repeated experiments was used for post-evaluation analysis). As exhibited in Figure 4 step 2, accuracy of 89.6 % to Hyp and 86.9% to Pro are obtained in the confusion matrix of post-evaluation analysis, which demonstrate the significant performance of the trained CNN model to differentiate the SERS spectra of AuNPs-NTA-Ni-Pro and AuNPs-NTA-Ni-Hyp.

To deeply analyze SERS spectra of Pro and Hyp, we used a statistical method to draw distribution histograms of the occurrence frequency for peaks assignments.30 The operation can extract specific features from fluctuated SERS spectra and reduce the influence of peak intensity, but average SERS spectra can’t achieve that. The histograms of the occurrence frequency for AuNPs-NTA (Figure S11) AuNPs-NTA-Ni-Pro (Figure 5a), AuNPs-NTA-Ni-Hyp (Figure 5b) at each peak position. AuNPs-NTA peaks frequency and assignment are seen in Figure S11, we took it as a reference for background removal. There is a band at 790 cm-1 of COOH deformation in AuNPs-NTA, a small shift to 793 cm-1 in AuNPs-NTA-Ni-Pro, and 798 cm-1 in AuNPs-NTA-Ni-Hyp.

Figure 5a showed the longest column bar at 485 cm-1 due to the Pro ring vibration, and the nearest peak at 455 cm-1 also from same vibration mode in AuNPs-NTA-Ni-Pro. Other Pro ring bending at 349, 380, 411, 618 cm-1, it shows an obvious fluctuation at 380 cm-1 from 362 to 392 cm-1, probably molecules were the fluctuated condition when NTA-Ni capturing Pro. The longer column bar at 825 cm-1 from the CH2 stretching of Pro ring but there is one short column is observed at 935 cm-1, suggesting Pro is likely to have different confirmation in this system due to the surface selection rules for SERS.31,32 1155 cm-1 displays middle column of CH2 wagging of Pro ring, and there is has a subtle wavenumber motion on 1233 cm-1 to 1247 cm-1 due to deformation with CH2 and NH2. These aromatic ring modes indicate Pro molecules entered the hot spot of AuNPs to get enhanced signals. Another higher column at 1290 cm-1 was assigned to bending of CH and NH group, they are the part moiety of Pro ring. These results mean that Pro has zwitterionic form in the experimental conditions with neutral solution.15 Especially, the band of nickel with oxygen was obtained at 286 cm-1, it demonstrated that nickel was attached to AuNPs-NTA and Pro was captured on AuNPs-NTA-Ni successfully.

In the same condition, Figure 5b gave Hyp information in AuNPs-NTA-Ni-Hyp system, the longest column at 405 cm-1 due to Hyp ring bending. Hyp has many common vibration modes with Pro due to their similar molecular structure. The strong bands at 487, 626, 936 cm-1 are associated with different Hyp ring deformation modes. The Hyp ring of CH2 stretching mode is clearly shown at 823 cm-1, thus making clear the presence of the protonated form of the Hyp.16 The strong band moves from 1210 to 1245 cm-1 belongs to Hyp ring deformation with CH2 and NH2, the higher column at 1288 cm-1 shows bending of CH and NH from ring, they are also consistent with a zwitterionic configuration of Hyp. It is reasonable to assume a configuration of AuNPs-NTA-Ni-Hyp that Hyp bending itself close to gold nanopillar, Hyp has an exceptional amino acid ring with tryptophan, so they have the similar mechanism of interaction between molecule and gold substrate in SERS measurement. Therefore, there are many significant SERS peaks form Hyp can easily collected. There is one band at 285 cm-1 associated with nickel and oxygen binding same as Pro. Particularly, Hyp has a strong band at 1169 cm-1, which is assigned to OH bending (seen in Figure 5b red marker). The results are combined with machine learning to prove that the AuNPs-NTA-Ni has good performance in distinguishing Hyp from Pro. More peaks assignment information in Table S2.

\section{Conclusion}
In summary, this platform be used to detect Pro and Hyp based on NTA-Ni specific capture ability. The time-series SERS spectra obtained many obvious and stable signals in some period, which gave more detailed information of Pro and Hyp. The distribution histograms of occurrence frequency provided the differences between NTA, Pro and Hyp, it achieved data fast and accurately analysis. There is high accuracy in the machine learning process, the test results of Pro and Hyp are 95.9% and 96.5%, the predict results of them are 86.9% and 89.9%, respectively. Importantly, we successfully found a new molecule (NTA-Ni) to specifically capture amino acids for the field of SERS biosensor. In future research, the strategy could be used for amino acids mixture analysis, NTA-Ni can capture every molecule with different absorption time to show different SERS spectrum. It will provide more possibilities for biomolecular detection, for example, rapid diagnosis for biomarkers of cancer diseases. 

\section{Supporting information}

Supplementary experiment section, Figures, and tables are in the supporting information. 
The codes for data analysis and all raw SERS data are in Github. The website link:
KuoZHAN/Surface-enhanced-Raman-spectroscopic-detection-of-proline-and-hydroxylated-proline-: CNN and post-evaluation model with raw data

\section{Acknowledgements}

This research receives support from Biocenter Oulu emerging project (DigiRaman), spreadhead project 2024 - 2027 and DigiHealth project (No. 326291), a strategic profiling project at the University of Oulu that is supported by the Academy of Finland and the University of Oulu. Yuan Zhang acknowledges China Scholarship Council for scholarship for doctoral study at the University of Oulu. We thank Professor Xia Yang (Southwest University) provided discussion about NTA.

\section{References}

(1)	Szabados, L.; Savoure, A. Trends Plant Sci. 2010, 15, 89–97.

(2)	Semenza,G.L. Annu. Rev. Cell. Dev. Biol. 1999, 15, 551–578.

(3)	Maxwell, P.H.; Wiesener, M.S.; Chang, G.W.; Clifford, S.C.; Vaux, E.C.; Cockman, M.E.; Wykoff, C.C.; Pugh, C.W.; Maher, E.R.; Ratcliffe, P.J. Nature 1999, 399, 271–275.

(4)	Bruick, R.K.; McKnight, S.L. Science 2001, 294, 1337–1340.

(5)	Epstein, A.C.; Gleadle, J.M.; McNeill, L.A.; Hewitson, K.S.; O'Rourke, J.; Mole, D.R.; Mukherji, M.; Metzen, E.; Wilson, M.I.; Dhanda, A.; Tian, Y.M.; Masson, N.; Hamilton, D.L.; Jaakkola, P.; Barstead, R.; Hodgkin, J.; Maxwell, P.H.; Pugh C.W.; Schofield C.J.; Ratcliffe P.J. Cell 2001, 107, 43–54.

(6)	Ivan, M.; Kondo, K.; Yang, H.; Kim, W.; Valiando, J.; Ohh, M.; Salic, A.; Asara, J.M.; Lane, W.S.; Kaelin, Jr.W.G. Science 2001, 292, 464–468.

(7)	Jaakkola, P.; Mole, D.R.; Tian, Y.M.; Wilson, M.I.; Gielbert, J.; Gaskell, S.J.; Kriegsheim, von. A.; Hebestreit, H.F.; Mukherji, M.; Schofield, C.J.; Maxwell, P.H.; Pugh, C.W.; Ratcliffe, P.J. Science 2001, 292, 468–472.

(8)	Zhou, T.; Erber, L.; Liu, B.; Gao, Y.K.; Ruan, H.B.; Chen, Y. Oncotarget 2016, 7, 79154–79169.

(9)	Srivastava, A.K.; Khare, P.; Nagar, H.K.; Raghuwanshi, N.; Srivastava, R. Curr. Protein Pept. Sci. 2016, 17, 596–602.

(10)	Robin J.M. Methods Mol. Med. 2005, 117, 189–207.

(11)	Ábrahám, E.; Hourton–Cabassa, C.; Erdei, L.; Szabados, L. Methods Mol. Biol. 2010, 639, 317–331.

(12)	Shen, T.T.; Zhang, C.; Liu, F.; Wang, W; Lu, Y.; Chen, R.Q.; He, Y. Sensors 2020, 20, 3229.

(13)	Cotte, J.F.; Casabianca, H.; Giroud, B.; Albert, M.; Lheritier, J.; Grenier–Loustalot, M.F. Anal. Bioanal. Chem. 2004, 378, 1342–1350.

(14)	Chan, S.W.P.; Greaves, J.; Da Silva, N.A.; Wang, S.W. BMC Biotechnol. 2012, 12, 51.

(15)	Cárcamo, J.J.; Aliaga, A.E.; Clavijo, E.; Garrido, C.; Gómez–Jeriaand, J. S.; Campos–Vallette, M.M. J. Raman Spectrosc. 2012, 43, 750–755.

(16)	Guerrero, A.R.; Aroca, R.F. J. Raman Spectrosc. 2012, 43, 478–481.

(17)	Yang, Z.C.; Chen, G.Q.; Shen, J.L.; Ma, C.Q.; Gu, J.; Zhu, C.; Li, L.; Gao, H. Spectrochim. Acta A Mol. Biomol. Spectrosc. 2023, 299, 122834.

(18)	Wang, K.F.; Zhang, S.Y.; Zhou, X.; Yang, X.; Li, X.Y.; Wang, Y.Q.; Fan, P.P.; Xiao, Y.Q.; Sun, W.; Zhang, P.K.; Li, W.F.; Huang, S. Nat. Methods 2024, 21, 92–101.

(19)	Wei, R.S.; Gatterdam, V.; Wieneke, R.; Tampe, R.; Rant, U. Nat. Nanotechnol. 2012, 11, 257–263.

(20)	Huang, J.A.; Zhao, Y.Q.; Zhang, X.J.; He, L.F.; Wong, T.L.; Chui, Y.S.; Zhang, W.J.; Lee, S.T. Nano Lett. 2013, 13, 5039–45.

(21)	Zhao, Y.; Huang, J.A.; Zhang, Z.; Chen, X.; Zhang, W. J. Mater. Chem. 2014, 2, 10218-10224. 

(22)	Wang, X.Y.; Liu, Q.H.; Tan, X.F.; Liu, L.Y.; Zhou, F.M. Analyst 2019, 144, 587–593.

(23)	Suzuki, T.; Takahashi, K.; Uehara, H.; Yamanobe, T. J. Therm. Anal. Calorim. 2013, 113, 1543–1549.

(24)	Huang, J.A.; Mousavi, M.Z.; Giovannini, G.; Zhao,Y.Q.; Hubarevich, A.; Soler, M.A.; Rocchia, W.; Garoli, D.; Angelis, F.D. Angew. Chem. Int. Ed. 2020, 59, 11423 – 11431.

(25)	Podstawka, E.; Ozaki, Y.; Proniewicz, L.M. Appl. Spectrosc. 2005, 59, 1516–1526.

(26)	Herlinger, A.W.; Long II, T.V. J. Am. Chem. Soc. 1970, 92, 6481–6486.

(27)	Hernández, B.; Coïc, Y.M.; Pflüger, F.; Kruglik, S.G.; Ghomi, M. J. Raman Spectrosc. 2016, 47, 210–220.

(28)	Rieley, H.; Kendall, G.K.; Zemicael, F.W.; Smith, T.L.; Yang, S. Langmuir 1998, 14, 5147–5153.

(29)	Shen, J.W.; Ahmed, T.; Vogt, A.; Wang, J.Y.; Severin, J.; Smith, R.; Dorwin, S.; Johnson, R.; Harlan, J.; Holzman, T. Anal. Biochem. 2005, 345, 258–269.

(30)	Li, W.; Guo, L.R.; Ding, X.L.; Ding, Y.R.; Ji, L.N.; Xia, X.H.; Wang, K. ACS Nano 2024, 18, 19200–19207.

(31)	Moskovits, M.; Suh, J.S. J. Phys. Chem. 1984, 88, 5526–5530.

(32)	Moskovits, M.; Suh, J.S. J. Phys. Chem. 1988, 92, 6327–6329.

\end{document}